\documentclass[12pt]{iopart}

\usepackage{graphicx}
\usepackage{url}
\begin{document}

\title[Modified electron trajectory due to axion dark matter background.]{Modified electron trajectory due to axion dark matter background}

\author{Yadir Garnica$^*$\footnote{Corresponding Author}$^1$,  and J Barranco$^1$}

\address{$^1$ Departamento de F\'isica, DCI, Campus Le\'on, Universidad de Guanajuato, Le\'on, 37150, }
\ead{\mailto{ya.garnicagarzon@ugto.mx}, \mailto{jbarranc@fisica.ugto.mx}}
\vspace{10pt}
\begin{indented}
\item[]October 2024
\end{indented}

\begin{abstract}
   It is well known that the coupling of the axion-like particle with the photon modifies Maxwell equations. One of the main consequences of these modifications is the conversion of axions into photons. Little has been said about other possible effects. In this note, we show that the trajectory of an electron can be significantly altered because of the emergence of an electric field by the axion-like particles dark matter background in this modified axion-electrodynamics. Different dark matter densities and magnetic field strengths are considered and it is shown that an axion-like particle with a mass $m_a\sim 10^{-22}$eV generates an electric field that can change the trajectory of an electron significantly in these scenarios.
\end{abstract}

\vspace{2pc}

%
%
%
\section{Introduction}
Just a few years after the Standard Model of elementary particles (SM) was established, the QCD axion was proposed to solve the strong CP problem \cite{Peccei:1977hh}. However, detection has not been done. Beyond the resolution of a fundamental problem in the SM, it was realized that the QCD axion could be a possible candidate to be the main component of dark matter (DM) \cite{Preskill:1982cy,Abbott:1982af,Dine:1982ah}. 
The QCD axion has the restrictive feature that the values of its mass $\mu$ and the decay constant $f_a$  are fixed by the confining group restriction such as 
\begin{equation}
\mu^2 f_a^2=m_{\pi}^2 f_\pi^2\,,\label{eq:mass_axion}
\end{equation}
where $f_\pi$ is the pion decay constant and $m_\pi$ the mass of the pion meson. This condition on the energy scale associated with the formation of condensates guarantees a nonzero vacuum expectation value. This condition severely restricts the possible values of $\mu$ to a certain group known as \textit{invisible axions}~\cite{kim:1979wk, bertolini:2015massive}.
Fifty years after its proposal, QCD-axion has not been detected. Severe limits on $\mu$ have been obtained; consequently, because of Eq.~(\ref{eq:mass_axion}), $f_a$ is also restricted \cite{Caputo:2024oqc,DiLuzio:2020wdo}.
Motivated highly for cosmological perspectives, the imminent proposal of a particle that shares all the interesting properties of the QCD axion without the severe restrictions for its mass given by Eq.~(\ref{eq:mass_axion}) had a relevant interest. These particles, which have the same phenomenological behavior as the QCD axion but do not intend to solve the strong CP problem, are called Axion-Like particles (ALP). An ALP could have a broad spectrum of masses, as low as $m_a \sim 10^{-22}~\mathrm{eV}$  \cite{Marsh:2015xka}, with a variety of energy decays. These ultralight scalar particles with no restriction on the mass and the decay constant could play the role of the dominant dark matter spectrum.
Among the different axion and ALP models, the prospect of indirectly measuring their existence presents a significant opportunity. Due to its "dark matter" features, the axion-photon coupling allows establishing boundaries over existing windows compared with the spectrum from microwaves to radio signals. This coupling implies modified electrodynamics~\footnote{It is possible to show the influence on the trajectory of a particle due to couplings with gauge bosons in different scenarios. Instead \cite{Visinelli:2024wyw}~explores how the transverse modes of an ultralight gauge field with dark matter features can modify neutrino propagation in galactic scenarios.}
 which has an explicit interaction term between the EM field and the axion field acting as an interaction source. 
This ALP-photon coupling has been explored at different scales looking for observational effects. From the atomic-size phenomena as changes in the Zeeman effect \cite{Sikivie:2014lha}, meter-size experiments looking for conversion from ALP to photons \cite{Sikivie:2013laa,millar2017:dielectric,Obata:2018vvr} or in Neutron Stars \cite{Barranco:2012ur, Iwazaki:2014wka,Raby:2016deh}.

In the following discussion, we will show that the same ALP-photon coupling can produce another observable effect: We will show that a charged particle immersed in a background of ALPs in the presence of an external magnetic field $\vec B$ will have a modified trajectory because the ALP-photon coupling induces an electric field. This effect can be relevant in extreme astrophysical scenarios. For instance, near a Supermassive black hole (SMBH) where the dark matter density is enhanced by a dark matter spike \cite{Gondolo:1999ef,Nampalliwar:2021tyz} and there exists a significant external magnetic field \cite{GRAVITY:2020hwn}. Likewise, the effect is relevant near a Neutron Star where the magnetic field can be as high as $10^{11}$T, even with moderate dark matter density. 
This emergent electric field has previously been computed for the case of the Earth \cite{IGRF-model}. In addition to the electric field, an oscillating magnetic field is generated near the surface of the Earth. In~\cite{IGRF-model}, by looking this emergent magnetic field in the magnetometer network data maintained by the SuperMAG Collaboration, competitive bounds on the ALP-photon coupling were obtained. Here, we will focus on the electric field instead and prove that if such field exist, some consequences in the trajectories of electrons should be induced.

 The organization of the paper is as follows. In section  \ref{modified_electro} we revisited the modifications to the Maxwell equations due to a possible coupling Axion-photon and derived the emergent electric field due to the existence of an axion background in the presence of an external magnetic field. In section \ref{discussion} the Lorentz-force equation for an electron trajectory is solved in various scenarios presented and discusses their possible physical detection. Some conclusions are presented in section \ref{conclusions}.

\section{Movement of a charged particle in an ALP background}\label{modified_electro}

\subsection{Axion electrodynamics and modified Maxwell equations}
The Lagrangian density for the QED $+$ axion has the general form (without the mass and the axion-fermion couplings)
\begin{equation}
   \mathcal{L}=
    -\frac{1}{4}F_{\mu\nu}F^{\mu\nu}-J^\mu A_\mu 
    +\frac{1}{2}\left(\partial_\mu a\partial^\mu a-m_a^2 a^2\right)
    -\frac{g_{a\gamma\gamma}}{4}a F_{\mu\nu}\tilde{F}^{\mu\nu}\, ,
    \label{eq:axion-lag-density}
\end{equation}
where the first two terms correspond to the EM strength tensor and the electric 4-current respectively, 
the second pair of terms are the kinetic and mass term for the axion field, 
and $\tilde{F}_{\mu\nu}$ is the dual tensor. The last term is the axion - photon coupling
with $g_{a\gamma\gamma}$ the coupling constant. For a general ALP, $g_{a\gamma\gamma}$ is given by
\begin{equation}
    g_{a\gamma\gamma}=\frac{\alpha_{EM}}{2\pi f_a}C_{a\gamma\gamma}\,
\, ,
\end{equation}
where $C_{a\gamma\gamma}$ is an effective coefficient that rescales in the function of the effective couplings \cite{d2021collider}. 
For the QCD axion particular case, we have an explicit relation between the strong phase transition temperature $\Lambda$, $m_a$ and $f_a$ as
\begin{eqnarray}
    \Lambda^4_{QCD}&=m_{a}^2 f_{a}^2\, , \qquad& \Lambda\sim 75\mathrm{MeV}\, ,
\label{eq:QCD-restr}
\end{eqnarray}
In the ALP context, the restrictions (\ref{eq:QCD-restr}) are not valid and $m_a, f_a$ are two independent parameters \cite{svrcek:2006axions, Arvanitaki:2010Str, conlon:2006qcd, bauer2021low}.

Minimizing the action Eq.~(\ref{eq:axion-lag-density}), the Euler-Lagrange equations associated correspond to the Maxwell equations with an additional modified term \cite{Wilczek:1987mv}
\numparts
\begin{eqnarray}
\bi{\nabla\cdot E}&=\rho-g_{a\gamma\gamma}\bi{B\cdot\nabla\cdot}a\, ,\\
    \bi{\nabla\times B}-\dot{\bi{E}}&=\bi{J}+g_{a\gamma\gamma}(\bi{B}\dot{a}-\bi{E\times\nabla}a)\, ,\\
    \bi{\nabla\cdot B}&=0\, ,\\
    \bi{\nabla\times E}+\dot{\bi{B}}&=0\, ,\\
    \ddot{a}-\bi{\nabla}^2 a+m_a^2 a &=g_{a\gamma\gamma} \bi{E\cdot B}\, .
\end{eqnarray}
\endnumparts
Now, to take into account the macroscopic density distribution of the different materials, we consider macroscopic charge and current densities, splitting them into a free and a bound component $\rho=\rho_f + \rho_b$ and $\bi{J}=\bi{J}_f+\bi{J}_b$, where the EM response is linked to the presence of the bound components. Using the macroscopic image through the polarization $\bi{P}$ and magnetization $\bi{M}$ vectors as
\begin{eqnarray*}
\rho_b&=-\bi{\nabla\cdot P}, \qquad & \bi{J}_b=\bi{\nabla\times M}+\dot{\bi{P}}\, .
\end{eqnarray*}
it is possible to define in the usual way the displacement  $\bi{D}$ and magnetization $\bi{H}$ vectors 
\begin{eqnarray}
\bi{D}&=\bi{E}+\bi{P}, \qquad& \bi{H}=\bi{B}-\bi{M}\, ,
\end{eqnarray}
and the free components satisfy the continuity equation
\begin{equation}
\dot{\rho}_f+\bi{\nabla}\cdot \bi{J}_f=0\, .
\label{eq:continuity}
\end{equation}

Then, the Maxwell-modified equations in the macroscopic frame have the form
\numparts
\begin{eqnarray}
\bi{\nabla\cdot D}&=\rho_f-g_{a\gamma\gamma}\bi{B\cdot\nabla\cdot}a\, ,\\
    \bi{\nabla\times H}-\dot{\bi{D}}&=\bi{J}_f+g_{a\gamma\gamma}(\bi{B}\dot{a}-\bi{E\times\nabla}a)\, ,\\
    \bi{\nabla\cdot B}&=0\, ,\\
    \bi{\nabla\times E}+\dot{\bi{B}}&=0\, ,\\
    \ddot{a}-\bi{\nabla}^2 a+m_a^2 a &=g_{a\gamma\gamma} \bi{E\cdot B}\, .
\end{eqnarray}
\endnumparts
Then, it is possible to simplify the system linearizing in the field components (we are interested in a $\bi{B}=\bi{B}_e$ static external field generated using an external current $\bi{J}_e$) and, keeping only the leading terms into $\bi{B}_e$, the non-homogeneous equations are
\numparts
\begin{eqnarray}
\bi{\nabla\cdot D}-\rho_f& = -g_{a\gamma\gamma}\bi{B}_e\cdot\bi{\nabla} a\, ,\label{eq:Gauss-modified}\\
\bi{\nabla\times H}-\dot{\bi{D}}-\bi{J}_f&=g_{a\gamma\gamma}\bi{B}_e\dot{a}\, ,\\
\ddot{a}-\bi{\nabla}^2 a+m_a^2 a&=g_{a\gamma\gamma}\bi{E}\cdot\bi{B}_e\, .
\end{eqnarray}
\endnumparts
The current $\bi{J}_e$ and the magnetization vector $\bi{M}_e$ satisfy their own Ampere-Maxwell equation, so the $\bi{J}_f$ vector is associated with the response of the material to the induced electric field. If we do not assume a static free charge density, we can write (\ref{eq:Gauss-modified}) (using (\ref{eq:continuity}) as
\begin{equation}
\bi{\nabla}\cdot(\dot{\bi{D}}-\bi{J}_f)=-g_{a\gamma\gamma}\bi{B}_e\cdot\bi{\nabla}\dot{a}\, .
\end{equation}
Expanding the fields in plane waves proportional to $e^{-i(\omega t-\bi{k}\cdot x)}$, we obtain
\numparts
\begin{eqnarray}
\bi{k}\cdot (\omega\hat{\bi{D}}+i\hat{\bi{J}}_f)&=-g_{a\gamma\gamma}\omega\bi{k}\cdot\bi{B}_e\hat{a}\, ,\\
\bi{k\times}\hat{\bi{H}}+\omega\hat{\bi{D}}+i\hat{\bi{J}}&=-g_{a\gamma\gamma}\omega\bi{B}_e\hat{a}\, ,\\
(\Omega_a^2-\bi{k}_a^2-m_a^2)\hat{a}&=-g_{a\gamma\gamma}\hat{\bi{E}}\cdot \bi{B}_e\, ,
\end{eqnarray}
\endnumparts
where $\hat{a}, \hat{\bi{D}}, \hat{\bi{H}}, \hat{\bi{E}}$ and $\hat{\bi{J}}_f$ are complex amplitudes depending on $\omega$ and $\bi{k}$. Assuming an isotropic and homogeneous material and a linear response, we can write the magnetic induction as $\hat{\bi{H}}=\hat{\bi{B}}/\mu$. If the medium is a conductor, the electric field drives a current given by the Ohm's law as $\hat{\bi{J}}_f=\sigma \hat{\bi{E}}$, where $\sigma$ is the material's conductivity, and, in the linear response domain, we can write $\hat{\bi{P}}=\chi\hat{\bi{E}}$, with $\chi$ being the susceptibility material \footnote{Here we follow the development presented in \cite{millar2017:dielectric}, but for our scenario, we are using $\bi{\rho}_f=\bi{J}_f=0$, thus $\epsilon\bi{E}=(1+\chi)\bi{D}$ without loss of generality.}. Therefore, we can accommodate the equation of the system using the effective dielectric permittivity~\cite{millar2017:dielectric} 
\begin{equation}
\epsilon=1+\chi+\frac{i\sigma}{\omega}\, ,
\end{equation}
getting the system
\numparts
\begin{eqnarray}
\epsilon \bi{k}\cdot\hat{\bi{E}}&=-g_{a\gamma\gamma}\bi{k}\cdot\bi{B}_e\hat{a}\, ,\label{eq:Gauss-modified1}\\
\bi{k}\times\hat{\bi{H}}+\omega\epsilon\hat{\bi{E}}&=-g_{a\gamma\gamma}\omega\bi{B}_e \hat{a}\, ,\label{eq:Ampere-modified}\\
\bi{k\cdot}\hat{\bi{B}}&=0\, ,\\
\bi{k\times}\hat{\bi{E}}-\omega\hat{B}&=0\, ,\\
(\Omega_a^2+\bi{k}_a^2-m_a^2)\hat{a}&=-g_{a\gamma\gamma}\hat{\bi{E}}\cdot\bi{B}_e\label{eq:axioneom}\, .
\end{eqnarray}
\endnumparts

In the case of axion or ALP dark matter, there is a field distribution $a(\vec x,t)$ that should be solved self-consistently in the system of Eqs.~(\ref{eq:Gauss-modified1}-\ref{eq:axioneom}). Next, we will restrict to the simplistic scenario where the field $a(\vec x,t)$ is constant. This is a crude approximation but for direct detection experiments, this is the most used zero-order approximation. Furthermore, we will focus on scales where the magnetic field is constant. In the case of the Earth, this approximation holds as long as the scale is on the order of kilometers. However, at scales comparable to the Earth's radius, the axion-ALP field will exhibit coordinate dependence, and the magnetic field will no longer remain constant. Therefore, a self-consistent solution is required. Nevertheless, in the next section, it will be shown that even under these coarse approximations some effects could be measurable. 

\subsection{A macroscopic electric field due to axion-photon coupling in the presence of an external magnetic field}

Then, in the homogeneous regime ($\bi{k}=0$), we can read directly the electric field induced by the axion presence from Eq.~(\ref{eq:Ampere-modified}) as
\begin{equation}
\bi{E}_a(t)=-\frac{g_{a\gamma\gamma}\bi{B}_e}{\epsilon}a(t)\,,
\end{equation}
thus a macroscopic electric field arises if there is an external magnetic field $\mathbf{B_e}$ and $a(t)\ne 0$.


If we assume that the DM is mainly composed of ALP and that the neighborhood of the Solar System is homogeneous, thus we can approximate the ALP field as
\begin{equation}
\Re(a(t))=\Re(a_0 e^{-i \Omega_a t})=a_0\cos(\Omega_a t)\label{eq:axion_dist}\, ,
\end{equation}
where in the case of interest, $a_0$ depends on the local DM distribution and 
\begin{equation}
 \Omega_a=1.5\times 10^9\left(\frac{m_a}{10^{-6}\mathrm{eV}}\right)\mathrm{Hz}\, .
\end{equation}
The local density of DM based on Milky Way stellar motions has been bounded around $\rho\in f_{DM}\left[0.3, 0.4\right]\mathrm{GeV ~cm^{-3}}$, where $f_{DM}\in\left[0,1\right]$ corresponds to the DM fraction made of ALP.

The ALP density is related to the field Eq.~(\ref{eq:axion_dist}) as
\begin{equation}
\rho_a=\frac{m_a^2\left|a_0\right|^2}{2}=f_{DM}\frac{300\mathrm{MeV}}{\mathrm{cm}^3}\, ,
\end{equation}
then, taking $5.1\times 10^{4}\mathrm{cm}^{-1}=1\mathrm{eV}$, we have
\begin{equation}
\rho_a=2.26\times 10^{-6}\mathrm{eV}^4\left(\frac{f_{DM}}{1}\right)\, ,
\end{equation}
or, solving for $a_0$ \cite{Graham:2011Axion}
\begin{equation}
a_0=\sqrt{\frac{2\rho_a}{m_a^2}}=2.13\times 10^3\mathrm{eV} \left(\frac{1\mu\mathrm{eV}}{m_a}\right)\sqrt{\frac{f_{DM} \rho_a}{300\mathrm{MeV/cm^3}}}\,.
\end{equation}
Therefore, we can write the induced electric field as
\begin{equation}
\bi{E}_a(t)=E_0 \cos\left(\Omega_a t\right)\frac{\bi{B_e}}{|\bi{B_e}|}\,, \label{Einduced}
\end{equation}
where
\begin{eqnarray}
    E_0&=\frac{g_{a\gamma\gamma}B_e}{\epsilon}a_0
    \approx 1.27\times 10^{-13}\frac{V}{m}\left|C_{a\gamma\gamma}\right|\left(\frac{5.7\times 10^{12}\mathrm{GeV}}{f_a}\right)
    \left(\frac{B_e}{1\mathrm{T}}\right)\times\nonumber\\
    &\times\left(\frac{1\mathrm{\mu eV}}{m_a}\right)\sqrt{\frac{\rho_a f_{dm}}{300\mathrm{MeV/cm^3}}}\, ,
\end{eqnarray}
for the last expression, we choose $\epsilon=\epsilon_0=1$.

\begin{figure}[t]  %
\centering
\includegraphics[scale=0.4]{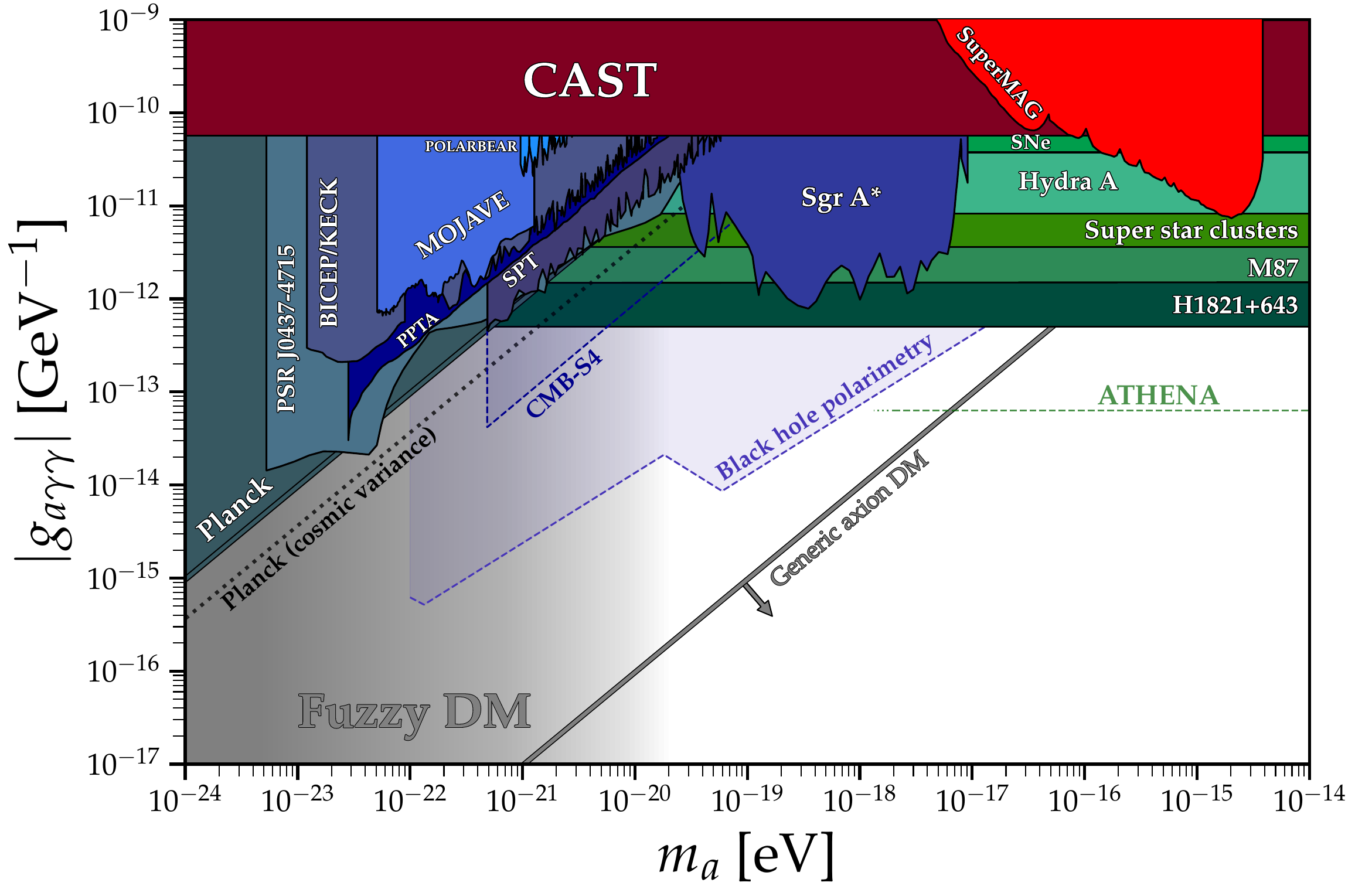}
\caption{$g_{a\gamma\gamma}$ limits and projections for ultralight ALP~\cite{AxionLimits}. }
\label{fig:ALP-limit-IGRF}
\end{figure}

$\bi{E}_a$ in Eq.~\ref{Einduced} is the simplest estimate of the induced electric field. Nevertheless, for the purpose of this work, that is, to show that this electric field can induce changes in the trajectories of charged particles in some astrophysical scenarios, we find this approximation suitable. 
For the case of the Earth, which poses a structured geomagnetic field, more detailed work is necessary. For the calculation of the induced electric field due to the presence of ultralight ALP dark matter particles and the presence of the geomagnetic field we will refer to \cite{ariel_arza_earth_2022}. The electric field  has the explicit form
\begin{equation}
    E_a^{IGRF}\sim g_{a\gamma\gamma}a_0(m_a L)^2 C_{\ell m}\, ,\label{E_IGRF}    
\end{equation}
where the $C_{\ell m}$ coefficients depends on the IGRF model~\cite{IGRF-model} and $L$ is the longest length scale. The particular geometry of the Earth as a perfectly conductive sphere near its surface surrounded by a vacuum gap and establishing no boundaries over the ionosphere and the magnetopause which surround the previous layer yields a screening effect where the induced electric field cannot be measured due to parametrically quadratic damping.
The electric field just above the Earth's surface is known to be $E_{Earth}=130$~N/C ~\cite{Rycroft:2000} . 
Demanding that the induced electric field $E_a^{IGRF}$ should be less than $E_{Earth}$, we cannot constrain the coupling constant $g_{a\gamma\gamma}$ effectively. In contrast to $\bi{E}_a$ for the constant magnetic field case, in the IGRF model, $E_a^{IGRF}$ is suppressed by a $(m_a L)^2$ factor, where $L$ is the measured distance varying from the Earth's radius to the ionosphere scale going to the magnetopause ($\sim 200R$). For the $C_{01}$ component, the restriction allowed for the $g_{a\gamma\gamma}$ due by demanding that $E_a^{IGRF}<E_{Earth}$  is outside the limits shown in the Fig.~\ref{fig:ALP-limit-IGRF} due to the restrictions related to the de Broglie's wavelength $L<\lambda_{DB}$. 

For the rest of this work, we will use the approximated induced electric field show in Eq. \ref{Einduced} to illustrate a possible effect on the movement of charged particles. 

\subsection{Lorentz force}
Once we have shown that there is an extra electric field induced by the ALP coupling with the external magnetic field $\bi{B_e}$ if we want to know the trajectory a charged particle will follow we need to solve the Lorentz force.
The velocity of the charged particle could be a relativistic one, thus we need to make relativistic corrections to the equation of motion (EoM). For this, we calculate the force correction as
\begin{eqnarray}
    F=\frac{d\vec{p}}{dt}=\frac{d}{dt}(m\vec{v})=m_{e}\frac{d}{dt}(\gamma \vec{v})=m_e\left(\gamma\frac{d\vec{v}}{dt}+\frac{d\gamma}{dt}\vec{v}\right)\, ,
\end{eqnarray}
where we use $m\rightarrow \gamma m_e$ where $m_e$ is the electron rest mass and as always $\gamma=\frac{1}{\sqrt{1-\frac{v^2}{c^2}}}$. The $\gamma$-derivative corresponds to~\footnote{Note that we will refer to the accelerations as $\vec a$, do not confuse with the axion field $a$.}
\begin{eqnarray}
    \frac{d\gamma}{dt}&=\frac{d}{dt}\left(\frac{1}{\sqrt{1-\frac{v^2}{c^2}}}\right)=\frac{\vec{v}\cdot \vec{a}}{c^2\left(1-\frac{v^2}{c^2}\right)^{3/2}}=\frac{\vec{v}\cdot \vec{a}}{c^2}\gamma^3\, .
\end{eqnarray}
Thus, the corrected expression for the total force is
\begin{equation}
    \vec{F}=m_e\left(\gamma\vec{a}+\gamma^3\frac{\vec{v
\cdot \vec{a}}}{c^2}\vec{v}\right)\, .
\label{eq:force}
\end{equation}
To calculate the charged particle trajectory, it is more useful to compute the acceleration vector (so we can split it into components and integrate it). Suggesting an ansatz $\vec{a}=A\vec{v}+B\vec{F}$ and putting it into the force expression
\begin{eqnarray}
    \frac{\vec{F}}{m_e\gamma}&=\frac{\gamma^2 \vec{v}}{c^2}\left(\vec{v}\cdot(A\vec{v}+B\vec{F})\right)+(A\vec{v}+b\vec{F})\\
    &=\frac{\gamma^2}{c^2}\vec{v}\left(A+B(\vec{v}\cdot\vec{F})\right)+(A\vec{v}+b\vec{F})\\
    &=\vec{v}\left(\frac{\gamma^2}{c^2}(A+B(\vec{v}\cdot \vec{F}))\right)+B\vec{F}\, .
\end{eqnarray}
Therefore, $B=\frac{1}{m_e\gamma}$ and
\begin{eqnarray}
    \frac{\gamma^2}{c^2}(A+B(\vec{v}\cdot \vec{F}))&=0, \qquad& \rightarrow A=-\frac{\vec{v}\cdot \vec{F}}{m_e\gamma}\, .
\end{eqnarray}
The acceleration vector has the final form
\begin{equation}
    \vec{a}=A\vec{v}+B\vec{F}=\frac{1}{m_e\gamma}\left(\vec{F}-\frac{(\vec{v}\cdot\vec{F})\vec{v}}{c^2}\right)\, ,
    \label{eq:acceleration}
\end{equation}
where we add the factor $1/c^2$ to the second term to adjust units.
In this case 
\begin{equation}
\vec{F}=\vec{F}_{Lorentz}=q\vec{E}+q\vec{v}\times\vec{B}\,,
\end{equation}
and the resulting EoM for a charged particle is
\begin{equation}
   \gamma m_e\vec{a}=q\vec{E}+q\vec{v}\times\vec{B}-\frac{q}{c^2}\left(\vec{v}\cdot\vec{E}\right)\vec{v}\,. \label{EoM}
\end{equation}
\section{Discussion}\label{discussion}
It is a textbook undergraduate exercise to solve the motion of 
an electron under the influence of an external magnetic field $\bi{B}_e=B_e \hat{z}$, with initial velocity components in the x-y plane. Our previous discussion here shows that if the Universe is immersed in an ALP background coupled with the photon field, thus a macroscopic electric field $\vec{E}_a$ emerges in the presence of an external magnetic field $\vec B$. We compute the trajectory of a electron from the Lorentz-force Eq.~(\ref{EoM}). 
For a constant external magnetic field $\vec{\bi{B}}=(0, 0, B_e)$ the induced electric field Eq.~(\ref{Einduced}) is 
\begin{equation}
\bi{E}_a(t)=E_0 \cos\left(\Omega_a t\right)\hat z\,. 
\end{equation}
With initial velocities $\vec{\bi{v}}_0=(v_{0x}, v_{0y}, 0)$,  the resulting equations of motion are
\numparts
\begin{eqnarray}
\frac{dv_x}{dt}&=\frac{q}{m_e \gamma}v_y B_e-\frac{1}{c^2 m_e \gamma}E_0 \cos(\Omega_a t)v_x v_z\, , \label{eq:dvxr}\\
\frac{dv_y}{dt}&=-\frac{q}{m_e \gamma}v_x B_e-\frac{1}{c^2 m_e \gamma}E_0 \cos(\Omega_a t)v_y v_z\, ,\label{eq:dvyr}\\
\frac{dv_z}{dt}&=-\frac{q}{m_e}E_0\cos(\Omega_a t)-\frac{1}{c^2 m_e \gamma}E_0 \cos(\Omega_a t) v_z^2\,.\label{eq:dvzr}
\end{eqnarray}
\endnumparts

\begin{figure}[t]  %
  \centering
   \includegraphics[scale=0.6]{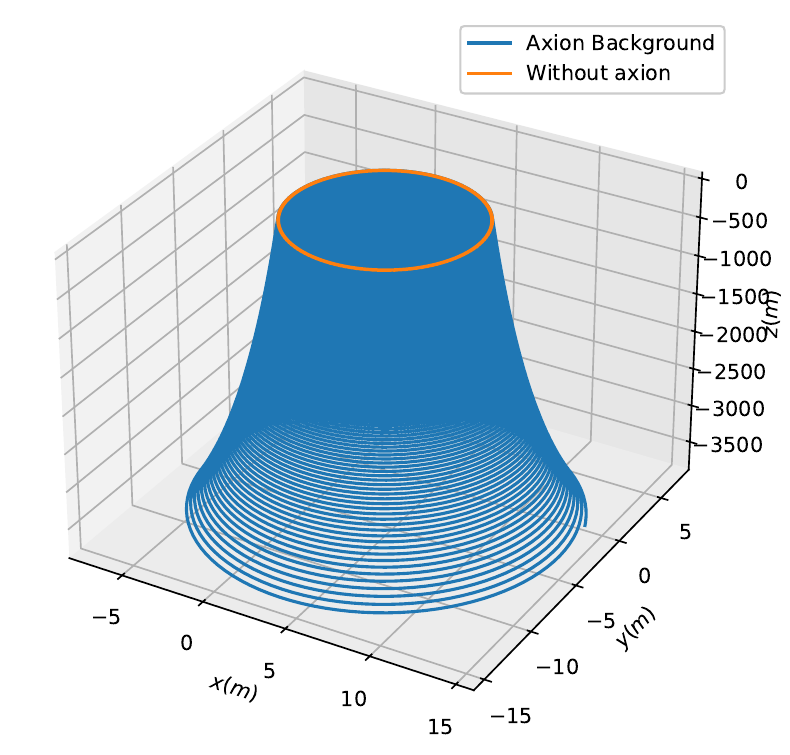}
  \caption{Trajectory of an electron immersed in an ALP background for $m_a=1\times 10^{-17}$eV, $f_a=10^{12}$GeV, $\rho_af_{DM}=0.3$GeV/cm$^{3}$, $C_{a\gamma\gamma}=1$, $B_e=0.5\times 10^{-4}$T. $v_{0x}=v_{0y}=\frac{1}{3}c$, $t_f=4\times 10^{-5}~\mathrm{s}$.}
  \label{fig:trajectory}
\end{figure}

A numerical solution is shown in Fig~\ref{fig:trajectory}. We considered typical values for the ALP particles $m_a=1\times 10^{-22}$eV, $f_a=10^{8}$GeV, $C_{a\gamma\gamma}=1$. The standard dark matter density at the Earth is $\rho_a =0.3f_{DM}~\mathrm{GeV/cm^{3}}$, and as a magnetic field, we took the magnetic field of the Earth $B_e=0.5\times 10^{-4}$T. As initial conditions we took $v_{0x}=v_{0y}=\frac{1}{3}c$ and we integrate form $t_i=0$ to $t_f=2\times 10^{-5}~\mathrm{s}$. We have included the trajectory that an electron must follow in the case where there are no axions or ALP particles. The modified trajectory is also shown and it can be observed that an appreciable deviation for the non-axion case.   

To have an estimator of the possible sizeable effects of the new electric field induced by ALPs in the electron trajectory, it is illustrative to see the exact solution in the non-relativistic case. 
Writing in components Eq.~(\ref{EoM}) with $\gamma \sim 1$ and neglecting the terms $v_x/c\ll 1$, $v_y/c\ll 1$ and $v_y/c\ll 1$, with the same geometry and conditions as we considered for the relativistic case, i.e. a constant external magnetic field $\vec{\bi{B}}=(0, 0, B_e)$, the equations of motion have the explicit form
\numparts
\begin{eqnarray}
\frac{dv_x}{dt}&=\frac{q}{m}v_y B_e, \label{eq:dvx}\\
 \frac{dv_y}{dt}&=-\frac{q}{m}v_x B_e,\label{eq:dvy}\\
 \frac{dv_z}{dt}&=-\frac{q}{m}E_0\cos(\Omega_a t)\,.\label{eq:dvz}
\end{eqnarray}
\endnumparts

The trajectory in the $x-y$ plane is obtained by integrating (\ref{eq:dvx})-(\ref{eq:dvy})
 \begin{eqnarray}
x(t)&=&\frac{v_{0x}}{\omega_e}\sin{\omega_e t}+x_0\,,\\
y(t)&=&\frac{v_{0x}}{\omega_e}\left(1-\cos{\omega_e t}\right)+v_{0y}t+y_0\,,
\end{eqnarray}
and taking $z(t=0)=0$, we get
\begin{equation}
z=z_a\left[\cos\left(1.51\times 10^9\left(\frac{m_a}{1\mu\mathrm{eV}}\right)\frac{t}{s}\right)-1\right]\, ,
\end{equation}
where
\begin{eqnarray}
    z_a&=9.9\times 10^{-21}\mathrm{m}~C_{a\gamma\gamma}\left(\frac{5.7\times 10^{12}\mathrm{GeV}}{f_a}\right)\left(\frac{B_e}{1\mathrm{T}}\right)\times\nonumber\\
    &\times\left(\frac{1\mathrm{\mu eV}}{m_a}\right)^3\sqrt{\frac{\rho_a f_{DM}}{300\mathrm{MeV/cm^3}}}\,.\label{eq:zeta}
\end{eqnarray}
and $\omega_e$ is related to a period of rotation as
\begin{equation}
T=\frac{2\pi}{\omega_e}=3.59\times 10^{-11}\mathrm{s}\left(\frac{1\mathrm{T}}{B_e}\right)\, .
\end{equation}

\begin{figure}[t]  %
\centering
\includegraphics[scale=0.6]{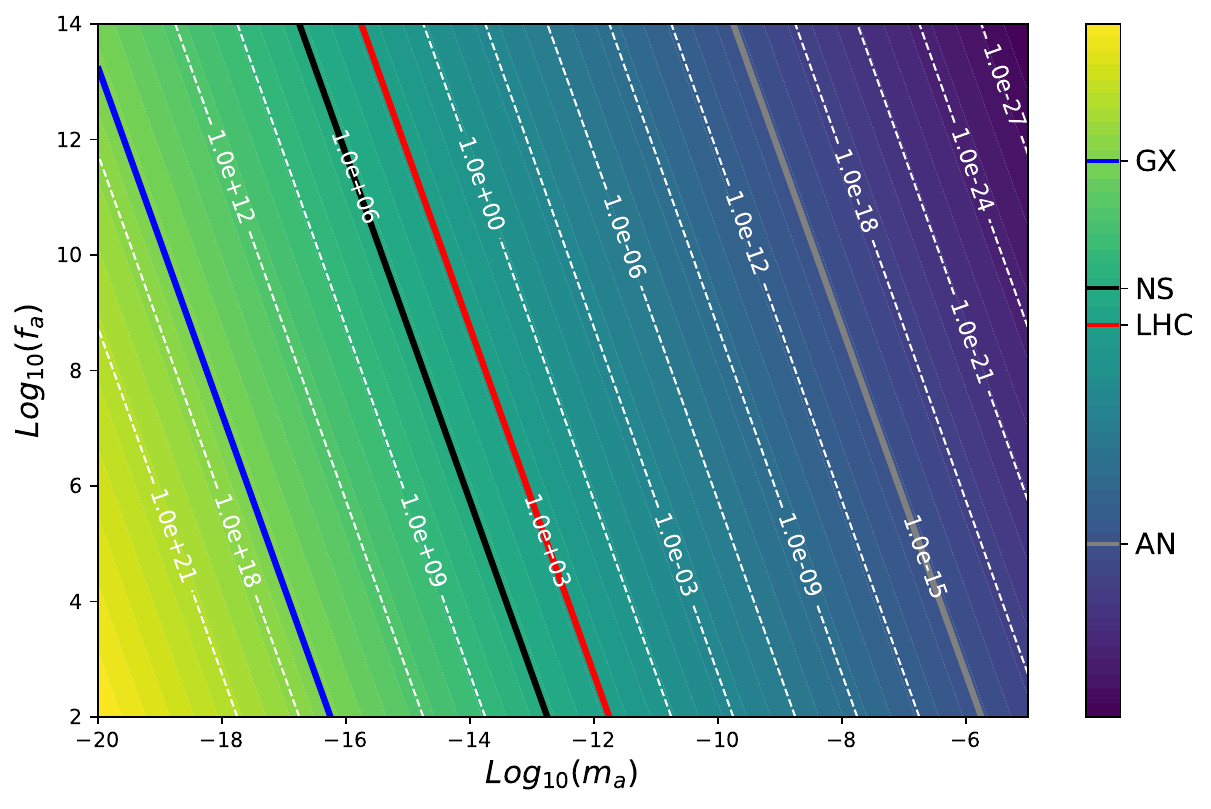}
\caption{Isocurves of $z_a$ in meters in the $m_a$, $f_a$ parameter space with reference values $\rho_a f_{DM}=300\mathrm{MeV/cm^3}$, $B_e=1$T and $C_{a\gamma\gamma}=1$. GX: Galaxy size, NS: Neutron stars, AN: Atomic Nucleus. }
\label{fig:earth}
\end{figure}
\begin{figure}[t]  %
  \centering
   \includegraphics[scale=0.6]{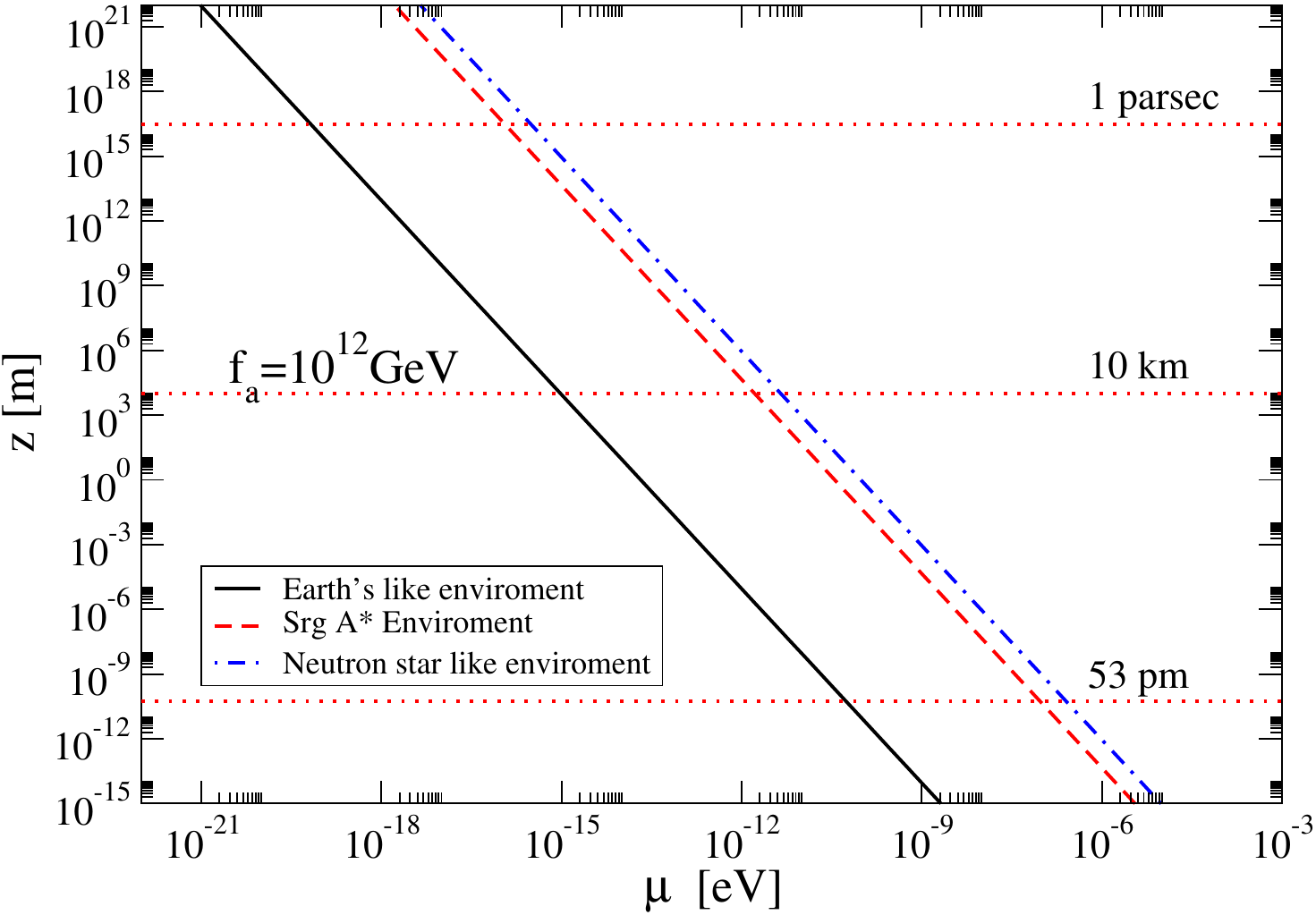}
  \caption{Possible changes in trajectory for different dark matter scenarios.}
  \label{fig:scenarios}
\end{figure}

The parameter $z_a$ in Eq.~(\ref{eq:zeta}) can help us in exploring different physical scenarios where this change in the trajectory of electrons (and of charged particles in general) is observed.
In Figure \ref{fig:earth} we show isocurves of $z_a(m_a,f_a)$ by fixing the dark matter density expected in the Solar System $\rho_a f_{DM}=300\mathrm{MeV/cm^3}$, the external magnetic field $B_e=1$T and $C_{a\gamma\gamma}=1$. We can see that $z_a$ can be as small as the typical size of atoms, or as large as the size of a galaxy. 
Extensive searches for QCD axion and ALP have been performed, most of them in the $\mu\mathrm{eV}$ region  \cite{ADMX:2018gho,ADMX:2019uok,Lee:2020cfj,HAYSTAC:2020kwv,Lee:2022mnc,Kim:2023vpo} and strong constraints on $f_a$ have been obtained. Astrophysical constraints are stronger and in general, for all masses of ALP, the current bound  is \cite{Viaux:2013lha,Ayala:2014pea,Sikivie:2020zpn}
\begin{equation}
f_a>10^{12}\mathrm{GeV}\,.\label{current_fa}
\end{equation}
By fixing $f_a=10^{12}$GeV, the isocurves in Figure \ref{fig:earth} can be analyzed in more detail as shown now in Figure \ref{fig:scenarios}. Solid black line represents a deviation of circular trajectories in an Earth-like situation, i.e. $\rho_a f_{DM}=300\mathrm{MeV/cm^3}$, and external magnetic field $B_e=0.5$~Gauss and $C_{a\gamma\gamma}=1$ with the current limit of $f_a$. From this, we conclude that deviation for axion-QCD particles will not affect an electron's trajectory. However, if we consider $m_a<10^{-15}$eV, the currently allowed values for $f_a$ still predict a sizeable effect on an electron's trajectory of the order of kilometers. Our analysis here does not include other effects that could as we are only solving the free electron trajectory.   
The searches of ALP and their possible effects are usually focused on places where the dark matter density could be high. Of special interest is the case of Supermassive black holes, where a high DM overdensity can be formed around it. This overdensity is known as the Dark matter spike \cite{Gondolo:1999ef}. For the case of our own SMBH in our own Galaxy, i.e. Sagittarius A$^*$ (Sgr A$^*$), the possible existence of a DM spike has been hypothesized and studied. Although this spike has not been observed, its density has been constrained, yet it could still be very high, namely, $\rho<10^{11}$ GeV/cm$^3$ \cite{Nampalliwar:2021tyz}. This value is a dozen order of magnitudes bigger than the expected dark matter density in the Solar System and thus, the possible change on a charged particle can be very high. This is shown in the red dotted line in Figure \ref{fig:scenarios} where we have computed $z_a$ for $\rho=10^{11}$GeV/cm$^3$ and we have fixed $B_e=100$G as inferred from astrophysical observations \cite{GRAVITY:2020hwn}. 
It is noteworthy that again for $m_a=10^{-15}$eV in the case of Sgr A$^*$ the deviation on the trajectory of an electron from the expected standard case but now under these conditions where ALP generates an extra electric field will be of the order of parsecs, that is, the relevant and possible measurable distances for the system under consideration.

Finally, in the same spirit of exploring different scenarios where the effect presented in this note can be observed, we consider now a regular dark matter density $\rho=0.3$ GeV/cm$^3$, but in this case huge magnetic fields as those observed in Magnetars where $B_e=10^{11}$T. This scenario is illustrated in Figure \ref{fig:scenarios} in a dotted blue line. For ALPs masses $m_a<10^{-12}$eV, the changes in the trajectory of the electron can be of the order of the system under study, that are Magnetars (with radius of the order of $10~\mathrm{km}$). 

\section{Conclusions}\label{conclusions}

In this note, we have studied the emergence of an electric field generated as a macroscopic effect of coupling of axions or ALP with the photon. In this case, an external magnetic field and the background of axion/ALP particles as dark matter are all over the universe. 
Furthermore, we have solved the equation of motion for a charged particle, specifically an electron. However, the same analysis can be done to any charged particle and we have observed that for the current allowed values of the free parameters for the axion-like particles, there is still the possibility that the electron will change severely its trajectory. We have briefly explored diverse astrophysical scenarios where this effect can be observed. 
The purpose here is not to constrain the free parameters of the ALP models, but to motivate the study of this effect in diverse physical situations where either we could detect indirectly the ALP particles or otherwise constrain the free parameters of the models.

\end{document}